\documentclass[10pt]{article}
\newcommand{\intsum}{\hspace{1mm}\int\hspace*{-4mm}\sum}

\usepackage{amssymb}
\usepackage{latexsym} 
\usepackage{mathptm}
\usepackage{epsfig}
\usepackage[dvips]{color}
\definecolor{gruen}{rgb}{0.3,0.0,0.2}
\definecolor{gelb}{rgb}{0.0,0.3,0.2}
\begin{document}

\markboth{SABINE HOSSENFELDER}{THE MINIMAL LENGTH AND LARGE EXTRA DIMENSIONS}

\title{THE MINIMAL LENGTH\\AND\\LARGE EXTRA DIMENSIONS}
\author{Sabine Hossenfelder\thanks{sabine@physics.arizona.edu}\\
\small{Department of Physics}\\ 
\small{University of Arizona}\\
\small{1118 East 4th Street}\\ 
\small{Tucson, AZ 85721, USA}}

\date{}
\maketitle


\begin{abstract}
Planck scale physics 
represents a future challenge, located between particle physics and general relativity. 
The Planck scale marks a threshold beyond which the old description of spacetime breaks
down and conceptually new phenomena must appear. Little is known about the fundamental
theory valid at Planckian energies, except that it necessarily seems to imply the
occurrence of a minimal length scale, providing a natural ultraviolet 
cutoff and a limit to the possible resolution of spacetime. 

Motivated by String Theory, the models of large extra dimensions lower the
Planck scale to values soon accessible.
These models predict a vast number of quantum gravity effects at the lowered
Planck scale, among them the production of TeV-mass 
black holes and gravitons.  
Within the extra dimensional scenario, also 
the minimal length comes into the reach of experiment and sets
a fundamental limit to short distance physics.

We review the status of Planck scale physics in these effective
models.
\end{abstract}

\section{The Planck Scale}	
Up to now, progress in physics has been marked by a progress in precision, experimentally as
well as theoretically, often the one going hand in hand with the other.
Physicists have extended the limits of their theories over and over again,  
included the principles of special relativity and quantum effects and formed
the Standard Model of particle physics which has proven its predictive power to extreme precision
and down to distances as small as $\approx 10^{-18}$~m. 

The question whether there exists a fundamental limit to the possibly achievable resolution of 
our spacetime is a reoccurring issue in science. 
It was in the 5th century b.c. that Demokrit postulated a smallest particle out of which matter is build. 
He called it an `atom'. In Greek, the prefix `a' means `not' and the word `tomos' means `cut'. 
Thus, atomos or atom means uncuttable or undividable. 
2500 years later, we know that not only the atom is dividable, but so is the atomic nucleus. 
The nucleus itself is a composite of neutrons and protons and further progress in science 
has revealed that even the neutrons and protons have a substructure. Is there an end to 
this or will the quarks and gluons turn out to be non-fundamental too? 
Any answer to this question has to include not only the structure of matter in our spacetime but
the structure of our spacetime itself -- it has to include gravity.

One of the major problem within the Standard Model is that the naive attempt to formulate 
a quantized theory of gravity fails. The length scale at which
the effects of quantum gravity are expected to become as important as the effects of
the other interactions is the so-called Planck length $l_{\rm p}$. Since gravity is so weak,
the Planck length is extremely small $l_{\rm p}\approx 10^{-35}$~m and effects could be
neglected in the experimentally accessible range so far. 

The Planck length which is derived from the Standard Model of physics is 
far out of reach for future experiments. 
But this depressing fact looks completely different if we work 
within the model of Large Extra Dimensions. Here, the energy scale of quantum gravity can be considerably
lowered -- which means, the minimal length will be raised and the quantum structure of spacetime might
lie only one order of magnitude beyond the current experimental possibilities.

Physics at Planckian energies
represents a challenge located between the interests of particle physics and
those of general relativity. At those energies, large gravitational fields are both generated and felt by
the interacting particles, which can then be used as sources and probes for quantum gravity effects. 
Planck scale physics is thus a key for our understanding of the fundamental laws of nature.

This brief review will focus on the phenomenological implications of Planck scale physics
and their examination in effective models. 

The outline of this review is as follows:
The next section serves as a motivation for the central subject of a minimal length. Section three will give an introduction into
the model with Large Extra Dimensions and its phenomenology. Section four will provide an
extension of this model which includes the minimal length. Observables of minimal length effects within
the extra dimensional scenario are discussed in section five, and we end with the conclusions in
section six. 

\section{The Minimal Length}
Gravity itself is inconsistent with physics
at very short scales.
The introduction of gravity into into quantum field theory appears to spoil their renormalizability and 
leads to incurable divergences. It has therefore been suggested that gravity
should lead to an effective cutoff in the ultraviolet, i.e. to a minimal observable length.
It is amazing enough that all attempts 
towards a fundamental theory of everything necessarily seem to imply the existence 
of such a minimal length scale. 

The occurrence of the minimal length scale has to be expected from very general reasons 
in all theories at high energies which attempt to include effects of
quantum gravity. This can be understood by a phenomenological argument which has been examined in various ways. 
Test particles of a sufficiently high energy to resolve a distance as small as the Planck length
are predicted to gravitationally curve and thereby to significantly disturb the very spacetime structure
which they are meant to probe. Thus, in addition to the expected quantum uncertainty, there is another
uncertainty caused which arises from spacetime fluctuations at the Planck scale.

Consider a test particle, described by a wave packet with a mean Compton-wavelength {$\lambda$}. 
Even in standard quantum mechanics, the particle suffers from an uncertainty in position $\Delta x$ 
and momentum $\Delta p$, given by the standard Heisenberg-uncertainty relation  $\Delta x \Delta p \geq 1/2$. 
Usually, every sample under investigation can be resolved by using beams of  
high enough momentum to focus the wave packet to a width below the size of the probe, $\Delta x$, which
we wish to resolve. 
The smaller the sample, the higher the energy must become and thus, the 
bigger the collider.

General Relativity tells us that a particle with an energy-momentum $p \sim 1/\lambda$ in a
volume of spacetime $\Delta x^3$ causes a fluctuation of the metric $g$. Einstein's Field
Equations allow us to relate the second derivative of the metric to the energy density
and so we can estimate the perturbation with\footnote{Here and throughout this review, we use the notation $\hbar=c=1$ which
leads to $m_{\rm p}=1/l_{\rm p}$ and $G=l_{\rm p}^2$.}
\begin{eqnarray}
\frac{\delta g}{\Delta x^2} \sim l_{\rm p}^2 \frac{p}{\Delta x^3} \quad.
\end{eqnarray}
This leads to an additional fluctuation, $\Delta x'$, of the spacetime coordinate frame of order 
$\Delta x' = \delta g \Delta x \approx l_{\rm p}^2 p$, which increases with momentum and becomes
non-negligible at Planckian energies. 

But from this expression we see not only that the fluctuations can no longer be neglected at
Planckian energies and the uncertainty of the measurement is amplified. In addition,
we see that a focussation of energy of the amount necessary to resolve the Planck scale
leads to the formation of a black hole whose horizon is located at $\delta g \approx 1$, should
the matter be located inside. Since the black hole radius has the property to expand linearly
with the energy inside the horizon, both arguments lead to the same conclusion: a minimal
uncertainty in measurement can not be erased by using test particles of higher energies. 
It is always  $\Delta x > l_{\rm p}$: at low energies because the Compton-wavelengths
are too big for a high resolution, at high energies because of strong curvature
effects.   

Thus, to first order the uncertainty principle 
is generalized to
\begin{eqnarray}
\Delta x \gtrsim \frac{1}{\Delta p} + \mbox{const.}~~ l_{\rm p}^2 {p}\quad. \label{gup}
\end{eqnarray}
Without doubt, the Planck scales marks a thresholds beyond which the old description of space-time 
breaks down and new phenomena have to appear.

More stringent motivations for the occurrence of a minimal length are manifold. 
A minimal length can be found in String Theory, Loop Quantum Gravity and Non-Commutative Geometries.
It can be derived from various studies of thought-experiments, from black hole physics, the holographic
principle and further more. Perhaps the most convincing argument, however, is that there seems to
be no self-consistent way to avoid the occurrence of a minimal length scale. For reviews see e.g. 
\cite{Garay:1994en,Ng:2003jk}.

To give a non-complete list:
\begin{enumerate}
\item String Theory naturally includes a minimal length scale as its very success arises from
the fact that interactions do no longer take place at one point in spacetime but are spread
out on the world sheet. By analyzing a spacetime picture of string scattering
at Planckian energies, is has been shown\cite{Gross:1987ar} 
that the extension of the string grows 
at high energies, 
leading to a generalized uncertainty relation of the form Eq. (\ref{gup}).\label{String}
\item Loop Quantum Gravity\cite{Rovelli:1997yv} is a non-perturbative approach to quantum gravity. Via the 
definition of so-called loop-states, the metric information is expressed in operators 
which in the classical limit yield the standard spacetime picture. In the quantum gravity 
regime, it turns out that some of these operators, e.g. 
the area operator, have a discrete spectrum which gives rise to a smallest-distance structure.\label{QLG}
\item Non-Commutative Geometries\cite{Szabo:2001kg} modify the algebra of the generators of 
space-time translations such that position measurements fail to commute. The commutation
relation is replaced with
\begin{eqnarray}
\left[x^i,x^j\right] = {\rm i} \theta^{ij} \quad,
\end{eqnarray}
where the tensor $\theta$ has a dimension of (length)$^{2}$ and measures the scale at which the
non-local effects become important. This modification can
be pursued throughout the usual development of quantum field theories, leading to a non-local
theory. It can be shown\cite{Douglas:2001ba} that within this approach a Gaussian
distribution naturally exhibits a maximally possible localization of width $\sim \theta$.\label{NGC}
\item A minimal length can be found from the Holographic Principle, which states\cite{'tHooft:1993gx} 
that the degrees of freedom of a 
spatial region are determined through the boundary of the region and 
that the number of degrees of freedom per Planck area is no greater 
than unity. This leads to a minimal possible uncertainty in length measurements\cite{Ng:2004bt}.\label{hol}
\item In the approach by Padmanabhan {\sl et al}\cite{Padmanabhan:1996ap}, 
the path integral amplitude is made invariant under a duality transformation
which replaces a length, $x$, by $x \to l_{\rm p}^2/x$. The proper distance
between two events in space-time, $\Delta x^2$, then is replaced by 
$\Delta x^2 \to l_{\rm p}^2 + \Delta x^2$, yielding a 'zero point length' of spacetime. 
Due to this, the divergences in quantum field theories are regularized by  removing the
small distance limit.\label{Pad}
\item Several phenomenological examinations of possible precision measurements\cite{Wigneretc}, 
thought experiments about 
black holes\cite{Scardigli:1999jh}
or the general structure of classical\cite{Daftardar:1986ci}, semi classical\cite{Kuo:1993if} and quantum-foamy 
space-time\cite{Ng:2004bt}. All of them lead to the conclusion that there exists 
a fundamental limit to distance measurement.
\end{enumerate}

The above listed points are cross-related in many ways. Point (\ref{NGC}) arises as a limit of
String Theory and also the imposed duality in (\ref{Pad}) is motivated by String T-duality. The
Holographic Principle has its origin in black hole physics and black hole physics itself connects to
almost every point mentioned before. Not listed are the topics of 
spacetime foams and  spin foams as they evade a brief description. The interested reader
is referred to \cite{Foam}.

Instead of finding evidence for the minimal scale as has been done in these numerous studies, on can 
use its existence as a postulate and derive extensions to quantum theories 
with the purpose to examine the arising properties.

Such approaches have been undergone and the analytical properties of the resulting theories have been 
investigated closely\cite{Kempf:1994su,minilengthgeneral,hydrogen}. 
In the scenario without extra dimensions, the derived modifications are 
important mainly for structure formation
and the early universe\cite{earlyuniverse}. The importance to deal with a finite resolution
of spacetime, however, is sensibly enhanced if we consider a spacetime with 
large extra dimensions.
 
\section{Large Extra Dimensions}
During the last decade, several models using compactified Large Extra Dimensions ({\sc LXD}s) as an
additional assumption to the quantum field theories of the Standard Model (SM) have
been proposed\cite{antoni}. The setup of these effective models is motivated by String Theory though
the question whether our spacetime has additional dimensions is well-founded on its own and
worth the effort of examination.

The models with {\sc LXD}s provide
us with an useful description to predict first effects beyond the SM.
They do by no means claim to be a theory of first principles or a candidate for a grand
unification. Instead, their simplified framework allows the derivation of 
testable results which can in turn help us to gain insights about the underlying theory.

There are different ways to build a model of extra dimensional space-time. Here, we want to
mention only the most common ones:
\begin{enumerate}
\item The {\sc ADD}-model proposed by Arkani-Hamed, Dimopoulos and Dvali\cite{add} adds $d$ extra
spacelike dimensions without curvature, in general each of them compactified to the same radius $R$. All 
SM particles are confined to our brane, while gravitons are allowed to propagate freely in the bulk. \label{ADD}
\item The setting of the model from Randall and Sundrum\cite{rs1,rs2} is a 5-dimensional spacetime with
an non-factorizable geometry. The solution for the metric is found by analyzing the solution of Einsteins 
field equations with an energy density on our brane, where the SM particles live. In the type I model\cite{rs1} 
the extra dimension is compactified, in the type II model\cite{rs2} it is infinite.
\item Within the model of universal extra dimensions\cite{uxds}
all particles (or in some extensions, only bosons) can propagate in the 
whole multi-dimen\-sional spacetime. The extra dimensions are compactified on an orbifold to 
reproduce SM gauge degrees of freedom.
\end{enumerate}
In the following we will focus on the model (\ref{ADD}) which yields a beautiful and simple explanation
of the hierarchy problem. Consider a particle of mass $m$ located in a spacetime of $d+3$ dimensions.
The general solution of Poissons equation yields its potential as a function of the radial distance $r$ to the
source
\begin{eqnarray}
V(r) \propto \frac{1}{M_{\rm f}^{d+2}} \frac{m}{r^{d+1}} \quad, \label{newtond}
\end{eqnarray}
where we have introduced a new fundamental mass-scale $M_{\rm f}$. The hierarchy problem then is the
question why, for $d=0$, this mass-scale is the Planck mass, $m_{\rm p}$, and by a factor $10^{16}$ smaller
than the mass-scales in the SM, e.g. the weak scale.

The additional $d$ spacetime dimensions are compactified on radii $R$, which are
small enough to have been unobserved so far. Then, at distances $r \gg R$, the extra dimensions
will 'freeze out' and the potential Eq. (\ref{newtond})
will turn into the common $1/r$ potential, but with a fore-factor given by the volume of the extra dimensions
\begin{eqnarray}
V(r) \to \frac{1}{M_{\rm f}^{d+2}} \frac{1}{R^{d}} \frac{m}{r} \quad. \label{newton} 
\end{eqnarray}
In this limit, we will rediscover the usual gravitational law which yields the relation
\begin{eqnarray}
m_{\rm p}^2 = M_{\rm f}^{d+2} R^d \quad. \label{Master}
\end{eqnarray}
Given that $M_{\rm f}$ has the right order of magnitude to be compatible with the other observed scales,
it can be seen from this argument that the volume of the extra dimensions suppresses the 
fundamental scale and thus, explains the huge value of the Planck mass.

The radius $R$ of these extra dimensions, for $M_{\rm f}\sim$~TeV, can be estimated with Eq.(\ref{Master}) and
typically lies in the range from mm to $10^3$~fm for $d$ from $2$ to $7$, or the inverse radius 
$1/R$ lies in energy range eV to MeV, respectively. The case $d=1$ is excluded. It would result in
an extra dimension about the size of the solar system. 

Due to the compactification, momenta in the direction of the {\sc LXD}s can only occur in quantized steps 
{$\propto 1/R$} for every particle which is allowed to enter the bulk. The fields can be expanded in
Fourier-Series
\begin{eqnarray}
\psi(x,y) = \sum_{n=-\infty}^{+\infty} \psi^{(n)}(x) \exp\left( {\rm i} n y/R \right) \quad,
\end{eqnarray} 
where $x$ are the coordinates on our brane and $y$ the coordinates of the {\sc LXD}s. 
This yields an infinite number of equally spaced excitations, the so called Kaluza-Klein-Tower.
On our brane, these massless KK-excitations act like massive particles, since the
momentum in the extra dimensions generates an apparent mass term
\begin{eqnarray}
\left[ \partial_x\partial^x - \left(\frac{n}{R}\right)^2 \right] \psi^{(n)}(x) = 0\quad.
\end{eqnarray} 

The most obvious experimental test for the existence of extra dimensions is a 
measurement of the Newtonian potential at sub-mm distances. Cavendish like experiments which search
for deviations from the $1/r$ potential have been performed during the last years with high
precision\cite{Newtonslaw}.

Also the consequences for high energy experiments are intriguing.
Since the masses of the KK-modes are so low, they get excited easily but it is not until
energies of order $M_{\rm f}$ that their phase-space makes them give an important contribution
in scattering processes.
The number of excitations $N(\sqrt{s})$ below a energy $\sqrt{s}$
\footnote{The estimated center of mass energy  for the LHC is $\sqrt{s}\approx$14~TeV.} 
can, for an almost
continuous spectrum, be estimated with the volume of the $d$-
dimensional sphere of radius $R\sqrt{s}$. We can then 
estimate the total cross-section for a point interaction, e.g. $e^+ e^- \to G \gamma$ ($G$ denotes
the graviton) by
\begin{eqnarray}
\sigma(e^+ e^- \to G \gamma) \approx \frac{\alpha}{m_{\rm p}^2}  N(\sqrt{s}) =  
\frac{\alpha}{s} \left( \frac{\sqrt{s}}{M_{\rm f}}\right)^{d+2} \quad,
\end{eqnarray} 
where we have used Eq.(\ref{Master}). As can be seen, at energy scales close to the new
fundamental scale the estimated cross-section becomes comparable to cross-sections of
electroweak processes.

The necessary Feynman rules for exact calculations of
the graviton tree-level interactions have be derived\cite{Hewett:2002hv} and the cross-sections have
been examined closely. 
Since the gravitons are not detected, their emission would lead to an energy loss in the 
collision and to a higher number of monojets. Modifications of SM predictions do
also arise by virtual graviton exchange, which gives additional contributions in the calculation
of cross-sections.

Another exiting signature of {\sc LXD}s is the possibility of black hole production. In the standard
$3+1$ dimensional space-time, the production of black holes requires a concentration of 
energy-density which can not be reached in the laboratory. As we have seen, 
in the higher dimensional space-time, gravity becomes stronger at small distances and
therefore the event horizon is located at a larger radius. 
We can estimate the horizon radius, $R_H$, of a mass $m$ by 
using the Newtonian potential Eq. (\ref{newtond}), where we will assume that the
black hole is small enough to completely fit into the extra dimensions $R_H\ll R$.
In the Newtonian limit, 
the radial entry of the metric tensor is approximately given by $g_{rr}=1- 2 V(r)$, and 
the horizon appears at the zero of $g_{rr}$ which leads to 
\begin{eqnarray}
R_H \sim \frac{1}{M_{\rm f}} \left(\frac{m}{M_{\rm f}}\right)^{\frac{1}{d+1}}  \quad.
\end{eqnarray} 
The exact formula which can be derived from the higher dimensional 
Schwarzschild-metric\cite{my} differs from the
given one by some numerical coefficients. It is not surprising to see that a black hole
with a mass about the new fundamental mass $m \approx M_{\rm f}$, has a radius of about the new fundamental
length scale $L_{\rm f}=1/M_{\rm f}$ (which justifies the use of the limit $R_H\ll R$). 
For $M_{\rm f}\sim$~1TeV this radius is $\sim 10^{-4}$~fm. Thus, at the {\sc LHC} it
would be possible to bring particles closer together than their horizon. A black hole could be created.
 
To compute the production details, the cross-section of the black holes can be approximated
by the classical geometric cross-section $\sigma\approx \pi R_H^2$. This  
cross section has been under debate\cite{Voloshin:2001fe}, 
but further investigations
justify the use of the classical limit at least up to energies of 
$\approx 10 M_{\rm f}$\cite{Solodukhin:2002ui} . However, the topic is still under
discussion, see also the very recent contributions\cite{Rychkov:2004sf}.

Setting $M_{\rm f}\sim 1$TeV and $d=2$ one finds 
$\sigma \approx 400$~pb.
With this, the total cross-section in hadronic processes can be computed and
shows that $\approx 10^9$ of these black holes per year  
could be produced at the {\sc LHC}\cite{dim}. 

Once produced, the black holes will undergo an evaporation process whose thermal
properties carry information about $M_{\rm f}$ and $d$. Furthermore,
crossing the threshold for black hole production causes a sharp cut-off for
high energetic jets as those jets now end up as black holes instead, and are re-distributed
into thermal particles of lower energies. Thus, black holes will give a clear signal.
For recent review on TeV-scale black holes see\cite{Kanti:2004nr}.

As we have seen, the {\sc LXD}-model predicts a rich phenomenology. Presently 
available data from collider physics as well as from astrophysics set 
constraints on the parameters of the model. For a recent update see 
e.g.\cite{Cheung:2004ab}.

\section{A Model for the Minimal Length}
The {\sc LXD} scenario predicts first effects of quantum gravity to 
appear close to the new fundamental scale $M_{\rm f}$. 
Signatures of gravitons and black holes 
would be observable in soon future at the {\sc LHC}.
For self-consistence, the model further must consider the fact that a lowering of the fundamental
scale leads to a raise of the minimal length. Within the model of {\sc LXD}s,
not only effects of quantum gravity occur at lowered energies but so do the effects of the minimal length scale.

To incorporate the notion of a minimal length into ordinary quantum field theory one can apply a
simple model which has been worked out in detail in\cite{Hossenfelder:2003jz,Hossenfelder:2004up}. 
Similar approaches have been used in \cite{Kempf:1994su,minilengthgeneral,hydrogen,earlyuniverse}.

\begin{figure}
\hspace*{-1.0cm}
\epsfig{figure=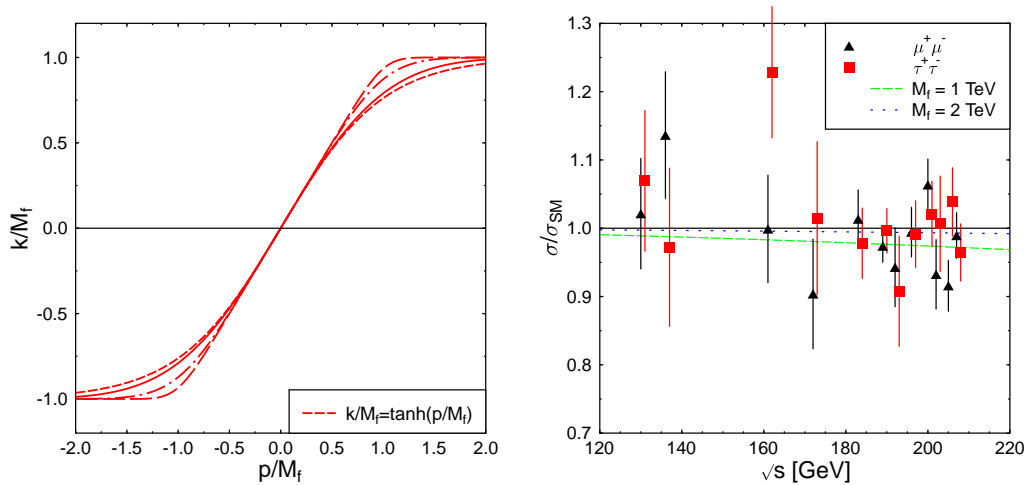, width=14cm}
\vspace*{-1.0cm}
\caption{The left plot shows various possible choices for the functional relation $k(p)$ used in the
literature. The long dashed line depicts the common choice 
$k= M_{\rm f}\tanh(p/M_{\rm f})$.
The right plot
\protect\cite{Hossenfelder:2003jz} 
shows the energy dependence of the ratio from the new 
total cross-section value to
the SM cross-section for fermionic pair 
production for different values of the minimal length. 
The data points are taken from \protect\cite{lep}.
 \label{fig1}}
\vspace*{-0.0cm}
\end{figure}

We assume, no matter how 
much we increase the momentum $p$ of a particle, its wavelength can never be decreased 
below some minimal length $L_{\mathrm f}$ or, equivalently, we can never increase
its wave vector $k$ above $M_{\mathrm f}$. Thus, the relation between the 
momentum $p$ and the wave vector $k$ is no longer linear, $p=k$, but a 
function.
\footnote{Note, that this is similar to introducing an energy 
dependence of Planck's constant $\hbar$.} 
This function $k(p)$ has to fulfill the following properties:
\begin{enumerate}
\item For energies much smaller than the new scale we reproduce the linear relation: 
for $p \ll M_{\mathrm f}$ we have $p \approx k$ \label{limitsmallp}.
\item It is an odd function (because of parity) and $k$ is collinear to $p$.
\item The function asymptotically approaches the upper bound $M_{\mathrm f}$. \label{upperbound}
\end{enumerate} 
An example for a function fulfilling the above properties is $k= M_{\rm f}\tanh(p/M_{\rm f})$, which
is a very common choice in the literature also known as the Unruh relation\cite{Unruh:1994je}. 
This function is depicted in Fig. \ref{fig1}, left, along with other possible choices.

In the absence of a first principle approach, these relations are chosen in an {\sl ad hoc} 
manner\footnote{Though the path integral approach by Padmanabhan\cite{Padmanabhan:1996ap} favours an 
particular relation.}. Results from such a model thus have to be examined as to their dependence 
on the functional
relation and their stability. However, the exact choice of the function is
important only in the regime $E\approx M_{\rm f}$ as at higher energies the same
asymptotic behavior must be fulfilled. In the regime of first effects, an expansion of the 
function can capture this ignorance in a free parameter. In the following, we will use the tanh as a concrete
example. An expansion to first order gives in this case
\begin{eqnarray}
k(p) &\approx& p - \frac{1}{3} p
\left( \frac{p}{M_{\rm f}} \right)^2 \quad. 
\end{eqnarray}

In the context of this model, it is further assumed that $L_{\rm f} \ll R$, such
that the spacing of the Kaluza-Klein excitations compared to energy scales $M_{\rm f}$ 
becomes almost continuous and we can use an integration instead of summing up the KK-tower.
  
The momentum $p$ is kept as the common Lorentz-invariant quantity which is conserved
in the standard (special relativistic) way. Since $k$ is a function of $p$ only,
the wave vector also is a constant of motion. In scattering processes 
with more than one participating particle, the wave vector $k$ in general 
will not be conserved as its transformation is non-linear.
  
In the presented approach, the mass-shell condition $E^2=p^2+m^2$ leads to a modified dispersion relation
$k(\omega)$ which is implicitly defined by $E(\omega)^2=p(k)^2+m^2$.  Inserting the
assumed relation $k(p)$ it can be put in
the general form 
\begin{eqnarray}
\omega^2 - k^2 - m^2 = \Pi(k,\omega) \quad. 
\end{eqnarray}
Thus, the investigation of models with modified dispersion relation is another way to
look at the minimal length proposal.

The quantization of the given ansatz is straightforward and follows the usual procedure. 
The commutators between the corresponding operators $\hat{k}$ and $\hat{x}$ 
remain in the standard form. 
Using the well known commutation relations and inserting the functional relation between the
wave vector and the momentum then yields the modified commutator for the momentum 
\begin{eqnarray}  
[\hat x_i,\hat k_j]={\mathrm i } \delta_{ij}\quad\Rightarrow\quad 
[\,\hat{x}_i,\hat{p}_j]&=& + {\rm i} \frac{\partial p_i}{\partial k_j} \quad.
\end{eqnarray}
This results in the generalized uncertainty principle ({\sc GUP)}
\begin{eqnarray} \label{gu}
\Delta p_i \Delta x_j \geq \frac{1}{2}  \Bigg| \left\langle \frac{\partial p_i}{\partial k_j} 
\right\rangle \Bigg| \quad, 
\end{eqnarray}
which reflects the fact that by construction it is not possible to resolve space-time distances
arbitrarily well. Since $k(p)$ gets asymptotically constant its derivative $\partial k/ \partial p$
drops to zero and the uncertainty in (\ref{gu}) increases for high energies. 
The behavior of our particles thus agrees with our expectations. 
Applying the approximation for the tanh one finds for the above relations the first order
alteration
\begin{eqnarray}
[\hat{x},\hat{p}] 
&\approx& {\rm i} \hbar\left( 1 + \frac{ \hat{p}^2}{M_{\rm f}^2} \right) \quad, \quad
\Delta p \Delta x \geq \frac{1}{2} \hbar \left( 1+ \frac{\langle \hat{p}^2 \rangle }{M_{\rm f}^2} \right) \quad.
\end{eqnarray}

Since $k=k(p)$ we have for the eigenvectors $\hat{p}(\hat{k})\vert k \rangle= p(k)\vert k \rangle$ and 
so $\vert k \rangle \propto \vert p(k) \rangle$. We could now add that both sets 
of eigenvectors have to be a complete orthonormal system and 
therefore $\langle k' \vert k \rangle = \delta(k-k')$, 
$\langle p' \vert p \rangle = \delta(p-p')$. 
This seems to be a reasonable choice at 
first sight, since  $\vert k \rangle$ is known from the low energy regime. 
Unfortunately, now the normalization of the states is different 
because $k$ is restricted to the Brillouin zone
$-1/L_{\mathrm f}$ to $1/L_{\mathrm f}$. 
To avoid the need to recalculate normalization factors, we  
choose the $\vert p(k) \rangle$ to be identical to 
the $\vert k \rangle$. Following the proposal of \cite{Kempf:1994su} this yields then
a modification of the measure in momentum space.
In the presence of $d$ {\sc LXD}s with radii $R$, the eigenfunctions are then normalized to\cite{Hossenfelder:2004up}
\begin{eqnarray} \label{norm}
\langle p'(k') \vert p(k) \rangle 
&=& (2 \pi )^{3+d}  \delta(p'_x-p_x) \Bigg| \frac{\partial p}{\partial k} \Bigg| \delta_{p_y'p_y}R^d
\quad,
\end{eqnarray}
where the expression in the vertical bar denotes the functional determinant of the relation $k(p)$ and
is responsible for an extra factor accompanying the
$\delta$-functions. The completeness relation of the modes takes the form
\begin{eqnarray} 
\intsum \frac{{\mathrm d}^3 k_x}{(2 \pi)^{d+3}} \frac{\langle k' \vert k \rangle}{N} = 
  R^d {\mathrm{Vol}}_d(p_y) 
\quad,
\end{eqnarray}
where ${\mathrm{Vol}}_d(p_y)$ denotes the volume of the $d$-dimensional momentum space. 
To avoid a new normalization $N$ of 
the eigenfunctions we take the factors into the integral by a redefinition of the measure in momentum space 
\begin{eqnarray} \label{rescalevolume3}
{{\mathrm d}^{d+3} k} \rightarrow {{\mathrm d}^{d+3} p}  \Bigg| 
\frac{\partial k}{\partial p} 
\Bigg| \frac{1}{{\mathrm{Vol}}_d(p_y) R^d} \quad.
\end{eqnarray}
This redefinition has a physical interpretation because we expect the momentum 
space to be squeezed at high momentum values and weighted less. 
With use of an expansion of the tanh for high energies we have  
$\partial k / \partial p  \approx \exp\left( - |p|/M_{\rm f}\right)$ 
and so we can draw the important conclusion that the momentum measure is exponentially squeezed at high energies.

One can now retrace the usual steps and derive equations of motion in quantum mechanics and quantum field theory.
It can be shown\cite{Hossenfelder:2003jz,Hossenfelder:2004up} that the general replacement $p\to p(k)$ 
and the new measure Eq. (\ref{rescalevolume3}) turn out
to be a very simple recipe to rewrite the usual equations and Feynman rules.

In quantum mechanics, the momentum operator in position representation can be derived 
and with this, the modified 
Schr\"odinger equation:  
\begin{eqnarray}  
\hat{\vec{p}} = -{\rm i} \; \hbar  {\nabla} \left(1 - L_{\rm f}^2 \Delta \right)
\quad,\quad
\hat{H} &=& 
- \frac{\hbar^2}{2m}    \left[ \nabla  \left(1 -  L_{\rm f}^2  \Delta \right)  \right]^2  + V(r) \quad. \label{schroed}
\end{eqnarray}
Further, the the quantization of the energy-momentum relation yields the modified Klein-Gordon Equation
and the Dirac Equation
\begin{eqnarray}  
\eta^{\mu\nu} 
\hat{p}_\nu(k) \hat{p}_\mu(k)  \psi =  m^2 \psi \quad,\quad (p\hspace{-1.6mm}/(k)-m)\psi&=&0  \quad,
\end{eqnarray}
where $p$ is now a function of $k$. 

\subsection{Lorentz Invariance}

The above given requirements for $k(p)$ do not list Lorentz-covariance.
It is easy to see that the proposed model can not conserve
this symmetry. Consider an observer who is, in the standard way, boosted relative to the minimal length.
He then would observe a contracted minimal length which would be even smaller than the minimal length. To
resolve this problem it is inevitable to modify the Lorentz-transformation. 
Several attempts to construct
such transformations have been made\cite{Amelino-Camelia:2002wr}, most notably the so-called 'Doubly Special
Relativity' by Amelino-Camelia.

In this approach, the modified Lorentz-transformations have two invariants. Not only is the speed of light, $c$,
a quantity independent of the reference frame, but so is the minimal length $L_{\rm f}$. Lorentz-invariance in this
setting is not broken but deformed at high energies. There is no preferred reference frame. The minimal
length is a fundamental limit for all observers.

Note, that the problem of finding the modified transformation between reference frames 
is identical to finding the  relation between $k$ and $p$. 
Since we assume that $p$ is a standard Lorentz vector, we will know how $k$ transforms if 
$k(p)$ is fixed (and the other way round).
Thus, the modified Lorenz-transformations which have gained an enormous amount of interest during the
last years are another way to look at the minimal length proposal.

\section{Observables}

The above derived properties of a quantum theory with a minimal length 
lead to modifications of the familiar equations in quantum mechanics and can be used to 
predict effects arising from the existence of a fundamental limit to space-time resolution.
 
\subsection{Collider physics}

With the modified version of the Feynman rules, it can be examined
how the minimal length influences SM cross-sections. A calculation
of cross-sections of {\sc QED} tree level processes {$A+B \to X+Y$} shows
that the result can be expressed in a particularly simple form.
The relation between the modified, $\tilde{\sigma}$, and the unmodified, 
$\sigma$, differential cross-section is given by
\begin{eqnarray}  
\frac{{\mathrm d} \tilde {\sigma}}{{\mathrm d} \sigma} = 
\prod_{n} \frac{E_n}{\omega_n}  
\Bigg|\frac{\partial k}{\partial p}\Bigg|_{ { {\bf p}}_{{\rm i}}={ {\bf p}}_{{\rm f}}} \quad,
\end{eqnarray}
where the index $n$ labels the participating particles and ${\bf p}_{\rm i}$
(${\bf p}_{\rm f}$) is is the three-momentum
of the initial (final) particles.
The dominant contribution to the cross-section 
is the squeezing-factor from the measure of momentum space which lowers the
phase space of the outgoing particles. This yields a cross-section
which is below the SM value. 

Applying this result to fermionic pair production processes 
{$e^+e^- \to \mu^+ \mu^-$}, or {$e^+e^- \to \tau^+ \tau^-$}, respectively, one finds that 
the modified cross-sections are in agreement with the data for a minimal length in the range 
{$L_{\rm f} \approx 10^{-4}$ fm}, see Fig \ref{fig1}, right. 

The effect of the squeezed momentum space does also {modify 
predictions from the {\sc LXD}-scenario}. The dilepton production under inclusion of virtual 
graviton exchange in hadron collisions 
has been examined in the minimal length scenario in ref.\cite{Bhattacharyya:2004dy}. It was
found that the bounds on the new fundamental scale $M_{\rm f}$ are less stringent than those
derived without a minimal length scale.

\begin{figure}
\hspace*{-1.0cm}
\epsfig{figure=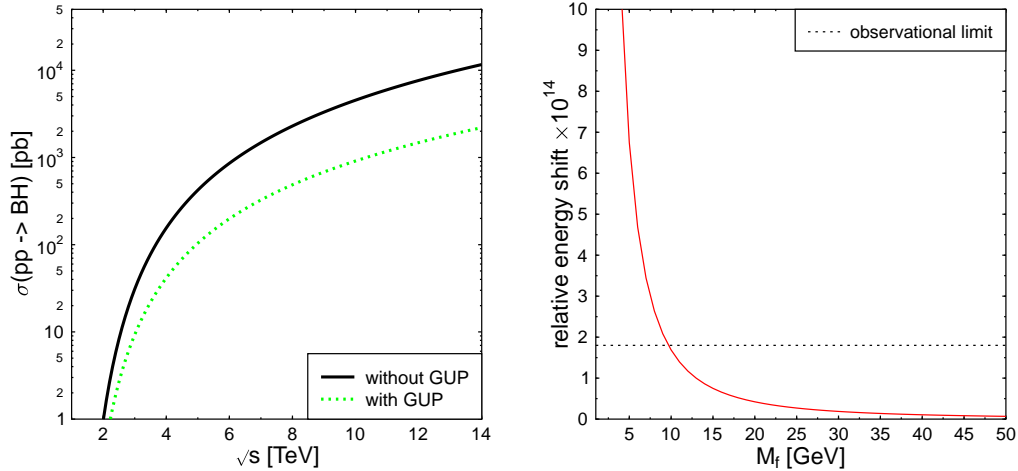, width=14cm}
\vspace*{-1.0cm}
\caption{The left plot\protect\cite{Hossenfelder:2004ze} shows the total cross-section 
for black hole production with minimal length $1/$TeV 
as a function of the center of mass energy $\sqrt{s}$. The ratio of the 
total cross section with and without {\sc GUP} for the expected 
{\sc LHC}-energy 
is $\approx 0.19$.
The right plot\protect\cite{Hossenfelder:2003jz} shows the relative shift of the S1-S2 energy level in the
hydrogen atom as a function of the minimal length. The horizontal line shows the current observational
accuracy\protect\cite{Niering} which is $\approx 1.8\times10^{-14}$. 
\label{fig2}
}
\end{figure}

Since the
modifications get important at energies close to the new scale, the predicted 
black hole production is 
strongly influenced. 
Black hole production is less probable \cite{Hossenfelder:2004ze} since the approach 
of the partons to 
distances below the Schwarzschild-radius requires higher energies within the minimal length scenario.
Figure \ref{fig2}, left, shows the total cross-section of the black hole production with and
without a generalized uncertainty principle. For the expected
{\sc LHC} energies  
the rate is lowered by a factor $\approx$~5.
 
The existence of a minimal length scale also influences 
the evaporation spectra of the produced black holes. This topic is
especially interesting with regard to the unsolved information loss problem which
arises if one applies the semi-classically derived result all the way down to
the Planck scale. In this case, 
the black hole emits thermal radiation whose sole property is its
temperature whatever the initial state of the collapsing matter has been. So, 
if the black hole
first captures all information behind its horizon and then completely vanishes into thermally
distributed particles, the basic principle of unitarity can be violated. This happens when 
the initial state was a pure quantum state and then evolves into a mixed one\cite{Hawk82}.

When one tries to escape the information loss problem there are two possibilities left: the information
is released back by a modification of the evaporation properties 
or a stable black hole remnant is left which keeps the
information. Once again, we see that black holes exist in a regime where 
quantum physics and gravity overlap and the interplay between such completely
different fields of physics as General Relativity, Thermodynamics and Quantum Theory 
shows us the limits of our current knowledge.

Using the generalized uncertainty principle, 
the properties\cite{Cavaglia:2004jw} of the evaporation process are modified when
the black hole radius approaches the minimal length
which decreases the number of decay particles and leads to the formation of
a stable remnant. 

\subsection{Astrophysics}

The investigation of Ultra High Energetic Cosmic Rays ({\sc UHECR}s) is a promising area for physics
beyond the SM as the observed 
spectrum extends to energies above $10^{20}$~eV\cite{Stecker:2003wm}. 

Theoretically, the {\sc UHECR}s, mostly protons, are expected to interact with the photons 
of the Cosmic Background Radiation ({\sc CBR}) to produce pions via the reactions  
$p + \gamma_{\rm CBR} \to n + \pi^0$, $p + \gamma_{\rm CBR} \to p + \pi^+$. 
The cross-section for this photopion production at the given center of mass energies 
is known from
experiments in the laboratory with high precision.
In the center of mass frame of the interacting particles,
the threshold for the photopion process is $\approx 1.08$~GeV\cite{Mucke:1999yb}. 
For the {\sc UHECR}s this causes the so-called Greisen-Zatsepin-Kuzmin ({\sc GZK}) 
cut-off\cite{Greisen:1966jv} at $E_{\rm GZK}\approx 5\times 10^{19}$~eV, above which
the universe becomes non-transparent for the traveling proton.
 
The highest energetic protons have most likely an extragalactic origin.
Given the known properties of the {\rm CBR}, the mean free path of the 
proton traveling towards the earth is much 
too low to allow protons from such distant sources to cause the observed {\sc UHECR}s.

To solve this {\sc GZK}-puzzle, it has been 
proposed that the absence of the cut-off is a consequence of a modified Lorentz-transformation 
coming along with the observer-independent minimal length scale\cite{UHECRs}. 

To see this, consider in which regard the traveling ultra 
high energetic proton differs from the one in the laboratory.
It travels through a soup of photons which looks pretty cold, $T_{\rm CBR}\approx 2.725$~K, 
in the earth rest frame and has a density of $\approx 400$~cm$^{-3}$. 
Under an ordinary Lorentz-transformation, however, this
soup will look very hot and dense for the high speed proton. The 
threshold for pion production is crossed and the mean free path will be 
lowered due to the increased density.

Now, the minimal length imposes a constraint on the number density of photons 
that the proton can experience. No matter how fast it travels and how much it is boosted,
the distance from one {\sc CBR} photon to the next will never be smaller than the minimal length.
Thus, compared to the standard case, 
its mean free path can be considerably raised, relaxing the {\sc GZK} cut-off. 

Note again, that Lorentz-invariance in this scenario is not broken but deformed. The apparent 
violation of
observer-independence is caused by a non-equivalence of reference frames. The reference frame
in which the {\sc CBR} is isotropic is exceptional and not identical with the rest frame of
the {\sc UHECR} proton.

\subsection{High Precision}
High precision measurements confirm the predictions of SM calculations with
impressing accuracy. Besides high energy physics, the detailed analysis at the
high precision frontier is likely to report first observed deviations from the SM. 
Current calculations of
SM processes include contributions of quantum field theory 
at higher loop order. Fortunately, such
effort is not necessary for an examination of the modifications arising
from a minimal length scale. Since the generalized uncertainty 
already influences quantum mechanical processes, an analysis of the
quantum mechanical equations of motion is sufficient to derive
constraints on the minimal length.

The derived modified Schr\"odinger equation Eq. (\ref{schroed}) will cause
a shift in the energy levels of atoms\cite{hydrogen}. The most precise data are available
for the Hydrogen S1-S2 transition. For the electromagnetic potential one 
inserts $V(r)=e^2/r$ and find a shift of the energy 
levels from the old values {$E_n$} to the new values $\widetilde{E}_n $
\begin{eqnarray}  
\widetilde{E}_n \approx E_n \left( 1 - \frac{4}{3} \frac{m_{\rm e}}{M_{\rm f}^2} \frac{E_0}{n^2} \right)\quad.
\end{eqnarray}
With the current accuracy of experiments for the Hydrogen S1-S2 transition, a very weak 
constraint on the new scale can be found $M_{\rm f} \gtrsim 10 {\mbox{ GeV}}$. The relative
shift as a function of $M_f$ is also depicted in Figure \ref{fig2}, right.
 
To derive more stringent predictions, an examination of the magnetic moment of the muon {$g_\mu-2$} 
\cite{Harbach:2003qz} can be done. The value of $g_\mu$ is modified by 
self energy corrections in quantum field theory.
Modifications from the minimal length are important even at the classical level and 
occur in the {\sc QED}-range.

A particles energy in a magnetic field {$B$} depends on its spin. 
Energy levels which are degenerated for free particles split up in the presence of a field.
In a static, homogeneous magnetic field, the expectation value of the spin vector will
perform a precession around the direction of the field. The rotation frequency is proportional to 
the strength of the field and the magnetic moment of the particle and so can be used to measure 
the magnetic moment.
The modifications arising from the minimal length can be derived by using Eq. (\ref{schroed})
and coupling the particle to the 
electromagnetic field {$K_\nu \to k_\nu - e A_\nu$} in the usual way    
\begin{eqnarray}   
\omega \vert\psi\rangle \approx \gamma^0\left( \gamma^i\hat{K}_i +\frac{m}{\hbar}\right)\left(1- \frac{\hbar^2 
\hat{K}^i\hat{K}_i + m^2}{M_{\rm f}^2}\right)\vert\psi\rangle \quad.
\end{eqnarray}
Analyzing this equation, one finds a constraint on the new scale in the range of the constraints 
from {\sc LXD}s: 
{$M_{\rm f} \approx 0.67$ TeV}. 
 
\subsection{Virtual processes}

One of the main motivations for the inclusion of a minimal length scale into the
framework of quantum field theory in extra dimensions, is the necessity
of an ultraviolet regulator for quantum gravity. In this regard, it is not surprising that
in the above discussed model with a minimal length scale, $L_{\rm f}$
acts as a natural regulator\cite{Hossenfelder:2004up}.
In contrast to the $4$-dimensional cases, in a higher dimensional space-time,
the emerging divergences can not be 
removed by applying an appropriate renormalization scheme 
(e.g. dimensional regularization) as every regularized result will depend 
on the regulator and must carefully be investigated as to its interpretation.
The minimal length therefore
reliably removes an inherent arbitrariness of choice of regulators in higher 
dimensional quantum 
field theories. 

To see this, consider a typical KK-loop amplitude with an external momenta restricted
to the brane $q=(q_x,0)$ in $d+4$ dimensions:
\begin{eqnarray}   
\Pi(q) \sim \sum_n \int  \frac{P(p)~~ {\rm d}^4 p }{\left[p_x^2+m_n^2\right]\left[(p-q)_x^2+m_n^2\right]} 
\quad, \label{loop}
\end{eqnarray}
where $m_n = (n/R)^2$ are the apparent masses of the KK-modes and $P(p)$ is a polynom
of order two whose precise (tensorial) structure is determined by the polarization tensor
of the gravitons\cite{Hewett:2002hv}. By taking the continuum limit
$\Delta m \to {\rm d}m$, Eq. (\ref{loop}) can be written as a $d+4$ dimensional integral
which is badly divergent
\begin{eqnarray}   
\Pi(q) \sim R^d \int  \frac{P(p)~~ {\rm d}^{4+d} p}{\left[p_x^2+p_y^2\right]\left[(p-q)_x^2+p_y^2\right]} 
\quad. \label{loop2}
\end{eqnarray}

Within the minimal length scenario, the one loop amplitude reads
\begin{eqnarray}   
\widetilde{\Pi}(q) \sim R^d \int   \Bigg| \frac{\partial k}{\partial p} \Bigg| 
\frac{P(p)~~ {\rm d}^{4+d} p }{\left[p_x^2+p_y^2\right]\left[(p-q)_x^2+p_y^2\right]} \quad, \label{loop3}
\end{eqnarray}
and is finite due to the exponentially squeezed phase space which appears via the
Jacobian determinant. Naturally, the result of this integration depends on the
regulator $L_{\rm f}$, but this quantity is no longer arbitrary and instead signals 
physically motivated corrections by the minimal length.

The so regularized loop-corrections have been used in Ref.\cite{Bhattacharyya:2004dy} for
the calculation of the virtual graviton contributions.

\section{Conclusions}
 
The occurrence of a minimal observable length scale arises from the attempt to include
quantum gravity into the Standard Model of particle physics. Should our spacetime turn
out to have additional large compactified dimensions, the 
effects of the minimal length would become testable soon.
Proposing a model to include a minimal length, we have shown that observable effects
have to be expected in high precision and high energy experiments. The predictions
from the extra dimensional scenario are modified, virtual contributions are naturally
regularized.
Furthermore, the existence of a minimal scale necessarily leads to a modification of
the Lorentz-transformations, which might explain the observed ultra-high energetic cosmic
ray events.

The central issue of the minimal length, however, remains the fundamental finiteness of
space-time resolution, which sets a limit to what future experiments possibly can achieve.
This limit might be reached soon.
 
\section*{Acknowledgments}

I thank Stefan Hofmann for clarifying the issue of Loop Quantum Gravity.
I would also 
like to thank K.~K.~Phua for the possibility to write this brief review. 

This work
was supported by a fellowship within the Postdoc-Programme of the German Academic 
Exchange Service 
({\sc DAAD}) and NSF PHY/0301998.

Due to the very finite length of this review, I apologize for every missing citation.

\end{document}